\documentclass[10pt, conference, compsocconf]{IEEEtran}
\usepackage{mathtools}

\usepackage{fancyvrb}
\usepackage{color}

\makeatletter
\def\PY@reset{\let\PY@it=\relax \let\PY@bf=\relax%
    \let\PY@ul=\relax \let\PY@tc=\relax%
    \let\PY@bc=\relax \let\PY@ff=\relax}
\def\PY@tok#1{\csname PY@tok@#1\endcsname}
\def\PY@toks#1+{\ifx\relax#1\empty\else%
    \PY@tok{#1}\expandafter\PY@toks\fi}
\def\PY@do#1{\PY@bc{\PY@tc{\PY@ul{%
    \PY@it{\PY@bf{\PY@ff{#1}}}}}}}
\def\PY#1#2{\PY@reset\PY@toks#1+\relax+\PY@do{#2}}

\expandafter\def\csname PY@tok@gd\endcsname{\def\PY@tc##1{\textcolor[rgb]{0.63,0.00,0.00}{##1}}}
\expandafter\def\csname PY@tok@gu\endcsname{\let\PY@bf=\textbf\def\PY@tc##1{\textcolor[rgb]{0.50,0.00,0.50}{##1}}}
\expandafter\def\csname PY@tok@gt\endcsname{\def\PY@tc##1{\textcolor[rgb]{0.00,0.27,0.87}{##1}}}
\expandafter\def\csname PY@tok@gs\endcsname{\let\PY@bf=\textbf}
\expandafter\def\csname PY@tok@gr\endcsname{\def\PY@tc##1{\textcolor[rgb]{1.00,0.00,0.00}{##1}}}
\expandafter\def\csname PY@tok@cm\endcsname{\let\PY@it=\textit\def\PY@tc##1{\textcolor[rgb]{0.25,0.50,0.50}{##1}}}
\expandafter\def\csname PY@tok@vg\endcsname{\def\PY@tc##1{\textcolor[rgb]{0.10,0.09,0.49}{##1}}}
\expandafter\def\csname PY@tok@vi\endcsname{\def\PY@tc##1{\textcolor[rgb]{0.10,0.09,0.49}{##1}}}
\expandafter\def\csname PY@tok@mh\endcsname{\def\PY@tc##1{\textcolor[rgb]{0.40,0.40,0.40}{##1}}}
\expandafter\def\csname PY@tok@cs\endcsname{\let\PY@it=\textit\def\PY@tc##1{\textcolor[rgb]{0.25,0.50,0.50}{##1}}}
\expandafter\def\csname PY@tok@ge\endcsname{\let\PY@it=\textit}
\expandafter\def\csname PY@tok@vc\endcsname{\def\PY@tc##1{\textcolor[rgb]{0.10,0.09,0.49}{##1}}}
\expandafter\def\csname PY@tok@il\endcsname{\def\PY@tc##1{\textcolor[rgb]{0.40,0.40,0.40}{##1}}}
\expandafter\def\csname PY@tok@go\endcsname{\def\PY@tc##1{\textcolor[rgb]{0.53,0.53,0.53}{##1}}}
\expandafter\def\csname PY@tok@cp\endcsname{\def\PY@tc##1{\textcolor[rgb]{0.74,0.48,0.00}{##1}}}
\expandafter\def\csname PY@tok@gi\endcsname{\def\PY@tc##1{\textcolor[rgb]{0.00,0.63,0.00}{##1}}}
\expandafter\def\csname PY@tok@gh\endcsname{\let\PY@bf=\textbf\def\PY@tc##1{\textcolor[rgb]{0.00,0.00,0.50}{##1}}}
\expandafter\def\csname PY@tok@ni\endcsname{\let\PY@bf=\textbf\def\PY@tc##1{\textcolor[rgb]{0.60,0.60,0.60}{##1}}}
\expandafter\def\csname PY@tok@nl\endcsname{\def\PY@tc##1{\textcolor[rgb]{0.63,0.63,0.00}{##1}}}
\expandafter\def\csname PY@tok@nn\endcsname{\let\PY@bf=\textbf\def\PY@tc##1{\textcolor[rgb]{0.00,0.00,1.00}{##1}}}
\expandafter\def\csname PY@tok@no\endcsname{\def\PY@tc##1{\textcolor[rgb]{0.53,0.00,0.00}{##1}}}
\expandafter\def\csname PY@tok@na\endcsname{\def\PY@tc##1{\textcolor[rgb]{0.49,0.56,0.16}{##1}}}
\expandafter\def\csname PY@tok@nb\endcsname{\def\PY@tc##1{\textcolor[rgb]{0.00,0.50,0.00}{##1}}}
\expandafter\def\csname PY@tok@nc\endcsname{\let\PY@bf=\textbf\def\PY@tc##1{\textcolor[rgb]{0.00,0.00,1.00}{##1}}}
\expandafter\def\csname PY@tok@nd\endcsname{\def\PY@tc##1{\textcolor[rgb]{0.67,0.13,1.00}{##1}}}
\expandafter\def\csname PY@tok@ne\endcsname{\let\PY@bf=\textbf\def\PY@tc##1{\textcolor[rgb]{0.82,0.25,0.23}{##1}}}
\expandafter\def\csname PY@tok@nf\endcsname{\def\PY@tc##1{\textcolor[rgb]{0.00,0.00,1.00}{##1}}}
\expandafter\def\csname PY@tok@si\endcsname{\let\PY@bf=\textbf\def\PY@tc##1{\textcolor[rgb]{0.73,0.40,0.53}{##1}}}
\expandafter\def\csname PY@tok@s2\endcsname{\def\PY@tc##1{\textcolor[rgb]{0.73,0.13,0.13}{##1}}}
\expandafter\def\csname PY@tok@nt\endcsname{\let\PY@bf=\textbf\def\PY@tc##1{\textcolor[rgb]{0.00,0.50,0.00}{##1}}}
\expandafter\def\csname PY@tok@nv\endcsname{\def\PY@tc##1{\textcolor[rgb]{0.10,0.09,0.49}{##1}}}
\expandafter\def\csname PY@tok@s1\endcsname{\def\PY@tc##1{\textcolor[rgb]{0.73,0.13,0.13}{##1}}}
\expandafter\def\csname PY@tok@ch\endcsname{\let\PY@it=\textit\def\PY@tc##1{\textcolor[rgb]{0.25,0.50,0.50}{##1}}}
\expandafter\def\csname PY@tok@m\endcsname{\def\PY@tc##1{\textcolor[rgb]{0.40,0.40,0.40}{##1}}}
\expandafter\def\csname PY@tok@gp\endcsname{\let\PY@bf=\textbf\def\PY@tc##1{\textcolor[rgb]{0.00,0.00,0.50}{##1}}}
\expandafter\def\csname PY@tok@sh\endcsname{\def\PY@tc##1{\textcolor[rgb]{0.73,0.13,0.13}{##1}}}
\expandafter\def\csname PY@tok@ow\endcsname{\let\PY@bf=\textbf\def\PY@tc##1{\textcolor[rgb]{0.67,0.13,1.00}{##1}}}
\expandafter\def\csname PY@tok@sx\endcsname{\def\PY@tc##1{\textcolor[rgb]{0.00,0.50,0.00}{##1}}}
\expandafter\def\csname PY@tok@bp\endcsname{\def\PY@tc##1{\textcolor[rgb]{0.00,0.50,0.00}{##1}}}
\expandafter\def\csname PY@tok@c1\endcsname{\let\PY@it=\textit\def\PY@tc##1{\textcolor[rgb]{0.25,0.50,0.50}{##1}}}
\expandafter\def\csname PY@tok@o\endcsname{\def\PY@tc##1{\textcolor[rgb]{0.40,0.40,0.40}{##1}}}
\expandafter\def\csname PY@tok@kc\endcsname{\let\PY@bf=\textbf\def\PY@tc##1{\textcolor[rgb]{0.00,0.50,0.00}{##1}}}
\expandafter\def\csname PY@tok@c\endcsname{\let\PY@it=\textit\def\PY@tc##1{\textcolor[rgb]{0.25,0.50,0.50}{##1}}}
\expandafter\def\csname PY@tok@mf\endcsname{\def\PY@tc##1{\textcolor[rgb]{0.40,0.40,0.40}{##1}}}
\expandafter\def\csname PY@tok@err\endcsname{\def\PY@bc##1{\setlength{\fboxsep}{0pt}\fcolorbox[rgb]{1.00,0.00,0.00}{1,1,1}{\strut ##1}}}
\expandafter\def\csname PY@tok@mb\endcsname{\def\PY@tc##1{\textcolor[rgb]{0.40,0.40,0.40}{##1}}}
\expandafter\def\csname PY@tok@ss\endcsname{\def\PY@tc##1{\textcolor[rgb]{0.10,0.09,0.49}{##1}}}
\expandafter\def\csname PY@tok@sr\endcsname{\def\PY@tc##1{\textcolor[rgb]{0.73,0.40,0.53}{##1}}}
\expandafter\def\csname PY@tok@mo\endcsname{\def\PY@tc##1{\textcolor[rgb]{0.40,0.40,0.40}{##1}}}
\expandafter\def\csname PY@tok@kd\endcsname{\let\PY@bf=\textbf\def\PY@tc##1{\textcolor[rgb]{0.00,0.50,0.00}{##1}}}
\expandafter\def\csname PY@tok@mi\endcsname{\def\PY@tc##1{\textcolor[rgb]{0.40,0.40,0.40}{##1}}}
\expandafter\def\csname PY@tok@kn\endcsname{\let\PY@bf=\textbf\def\PY@tc##1{\textcolor[rgb]{0.00,0.50,0.00}{##1}}}
\expandafter\def\csname PY@tok@cpf\endcsname{\let\PY@it=\textit\def\PY@tc##1{\textcolor[rgb]{0.25,0.50,0.50}{##1}}}
\expandafter\def\csname PY@tok@kr\endcsname{\let\PY@bf=\textbf\def\PY@tc##1{\textcolor[rgb]{0.00,0.50,0.00}{##1}}}
\expandafter\def\csname PY@tok@s\endcsname{\def\PY@tc##1{\textcolor[rgb]{0.73,0.13,0.13}{##1}}}
\expandafter\def\csname PY@tok@kp\endcsname{\def\PY@tc##1{\textcolor[rgb]{0.00,0.50,0.00}{##1}}}
\expandafter\def\csname PY@tok@w\endcsname{\def\PY@tc##1{\textcolor[rgb]{0.73,0.73,0.73}{##1}}}
\expandafter\def\csname PY@tok@kt\endcsname{\def\PY@tc##1{\textcolor[rgb]{0.69,0.00,0.25}{##1}}}
\expandafter\def\csname PY@tok@sc\endcsname{\def\PY@tc##1{\textcolor[rgb]{0.73,0.13,0.13}{##1}}}
\expandafter\def\csname PY@tok@sb\endcsname{\def\PY@tc##1{\textcolor[rgb]{0.73,0.13,0.13}{##1}}}
\expandafter\def\csname PY@tok@k\endcsname{\let\PY@bf=\textbf\def\PY@tc##1{\textcolor[rgb]{0.00,0.50,0.00}{##1}}}
\expandafter\def\csname PY@tok@se\endcsname{\let\PY@bf=\textbf\def\PY@tc##1{\textcolor[rgb]{0.73,0.40,0.13}{##1}}}
\expandafter\def\csname PY@tok@sd\endcsname{\let\PY@it=\textit\def\PY@tc##1{\textcolor[rgb]{0.73,0.13,0.13}{##1}}}

\def\PYZus{\char`\_}
\def\PYZob{\char`\{}
\def\PYZcb{\char`\}}

\def\PYZlt{\char`\<}
\def\PYZgt{\char`\>}

\def\PYZhy{\char`\-}

\makeatother

\ifCLASSINFOpdf
  \usepackage[pdftex]{graphicx}
\else
  \usepackage[dvips]{graphicx}
\fi
\usepackage{algorithmic}

\usepackage{flushend}

\begin{document}
%
\title{Fast Access to Columnar, Hierarchically Nested Data via Code Transformation}


\author{\IEEEauthorblockN{Jim Pivarski, Peter Elmer}
\IEEEauthorblockA{Physics Department\\
Princeton University\\
Princeton, NJ, 08544\\
pivarski@princeton.edu, peter.elmer@cern.ch}
\and
\IEEEauthorblockN{Brian Bockelman, Zhe Zhang}
\IEEEauthorblockA{Computer Science and Engineering\\
University of Nebraska-Lincoln\\
Lincoln, NE 68588\\
bbockelm@cse.unl.edu, zhan0915@huskers.unl.edu}
}


%


\maketitle

\begin{abstract}
Big Data query systems represent data in a columnar format for fast, selective access, and in some cases (e.g.\ Apache Drill), perform calculations directly on the columnar data without row materialization, avoiding runtime costs.

However, many analysis procedures cannot be easily or efficiently expressed as SQL. In High Energy Physics, the majority of data processing requires nested loops with complex dependencies. When faced with tasks like these, the conventional approach is to convert the columnar data back into an object form, usually with a performance price.

This paper describes a new technique to transform procedural code so that it operates on hierarchically nested, columnar data natively, without row materialization. It can be viewed as a compiler pass on the typed abstract syntax tree, rewriting references to objects as columnar array lookups.

We will also present performance comparisons between transformed code and conventional object-oriented code in a High Energy Physics context.

\end{abstract}


%
\IEEEpeerreviewmaketitle

\section{Motivation}

Data analysts in the field of High Energy Physics (HEP) routinely deal with terabyte and petabyte scale datasets, but access them as objects persisted in files, rather than databases. Thus, they miss out on advantages enjoyed by data analysts in other fields, such as automated scale-out, data replication, primary key indexing for faster selections, and term rewriting and query planning of high-level queries.

A major reason for this is that HEP data do not fit any of the existing database models, including common non-relational (NoSQL) ones: HEP data are hierarchically nested with arbitrary-length collections within collections, not the rectangular table served well by SQL. This feature suggests a document store like MongoDB\cite{mongodb}, except that HEP data have regular structure that is sparsely accessed: only a few of the object attributes are needed in each query, which would waste disk access if all attributes of an object are stored contiguously, rather than in ``columns,'' in which all values of a given attribute, across objects, are contiguous on disk. Since HEP objects typically have hundreds of attributes, accessing a few of them in a typed columnar store is orders of magnitude faster than a schemaless document store. Unlike key-value tables, HEP data must be scanned in bulk; unlike graph databases, HEP data cross-linking is limited to small, disconnected graphs called ``events'' no larger than hundreds of kilobytes.

Furthermore, most of the well-known database systems use SQL or an SQL variant as the query language, but even the simplest HEP analysis functions are awkward and possibly hard to optimize as SQL. HEP analysis functions typically iterate over combinations of objects in different subcollections within each event, which would require multiple SQL explodes and joins. Although the hierarchically nested structure can be described in modern SQL,

\vspace{0.15 cm}

\begin{Verbatim}[commandchars=\\\{\}]
\PY{k}{CREATE} \PY{k}{TYPE} \PY{n}{PARTICLE} \PY{k}{FROM}
    \PY{n}{STRUCT}\PY{o}{\PYZlt{}}\PY{n}{pt}\PY{p}{:} \PY{n}{DOUBLE}\PY{p}{,}
           \PY{n}{eta}\PY{p}{:} \PY{n}{DOUBLE}\PY{p}{,}
           \PY{n}{phi}\PY{p}{:} \PY{n}{DOUBLE}\PY{o}{\PYZgt{}}\PY{p}{;}

\PY{k}{CREATE} \PY{k}{TABLE} \PY{n}{events} \PY{p}{(}
    \PY{n}{eventId} \PY{n+nb}{INT}\PY{p}{,}
    \PY{n}{muons}   \PY{n+nb}{ARRAY}\PY{o}{\PYZlt{}}\PY{n}{PARTICLE}\PY{o}{\PYZgt{}}\PY{p}{,}
    \PY{n}{jets}    \PY{n+nb}{ARRAY}\PY{o}{\PYZlt{}}\PY{n}{PARTICLE}\PY{o}{\PYZgt{}}\PY{p}{,}
    \PY{k}{UNIQUE} \PY{k}{KEY} \PY{n}{eventId}
\PY{p}{)}\PY{p}{;}
\end{Verbatim}

\vspace{0.15 cm}
\noindent often the first exploratory action in a HEP analysis is to plot a distribution of the highest particle {\tt pt} {\it per event}. In SQL, one would have to explode the {\tt muons} into a virtual table and aggregate for maximum {\tt pt}, grouped by {\tt eventId}, but the database system should be made aware that the millions of events are small and should each remain local to avoid millions of collations over the network. To search for short-lived particles that might have decayed into one muon and one jet, the analyst would have to iterate over all possible pairs of {\tt muons} and {\tt jets}, computing the mass\footnote{$\sqrt{2 {p_1}^{p_T} {p_2}^{p_T} (\cosh({p_1}^\eta - {p_2}^\eta) - \cos({p_1}^\phi - {p_2}^\phi))}$ for $p_1$ and $p_2$} of each combination, with or without constraining to one candidate per event. Searches for particles decaying into two muons must avoid double-counting unordered muon pairs, etc. In general, exploding subcollections into virtual tables and joining on {\tt eventId} is both syntactically inconvenient for the data analyst and introduces an optimization problem for the database engineer. A short functional or procedural program applied to each event is much more natural than shoehorning the problem into SQL.

However, even query language agnostic systems like Apache Drill\cite{drill} make SQL-motivated assumptions about the structure of queries in the query planning and distribution. Apache Spark\cite{spark} drops from efficient SparkSQL\cite{sparksql} processing to a slower mode when the user needs to apply an arbitrary function. This is unnecessary. The features that accelerate scans over tables, namely a columnar data layout, Just-In-Time (JIT) compilation, and avoiding row materialization, can be applied to generic programming languages.

This paper describes one component of a database/fast query system for HEP data, which is in early development. Such a system will involve distributed processing, Hadoop-style data locality, indexing, and data management with columnar, rather than file, granularity. However, this paper focuses only on the execution engine, which performs JIT compilation without object materialization on hierarchically nested data structures stored in a columnar layout.

JIT compilation is central to machine learning tools like H2O\cite{h2o} and Theano\cite{theano}, as well as Julia\cite{julia}, a scientific programming language, and ROOT\cite{root}, the analysis framework that is ubiquitous in HEP. However, these tools do not avoid object materialization, even if the data are persisted in a columnar layout, as in the case of ROOT. Most HEP data are currently stored as ROOT files, which represent hierarchically nested objects in columnar form, yet the ROOT framework materializes them as C++ objects before applying analysis functions. We have augmented ROOT\cite{bulkio} to avoid this object materialization and modify the user's analysis function instead. Leaving the data in columnar arrays, we walk over an Abstract Syntax Tree (AST) of the user's analysis function, replacing object references with array element retrievals, then pass the transformed function to a traditional compiler. For the analyst's convenience, we transform functions written in Python and compile the result with Numba\cite{numba}, which allows high-level querying yet produces bytecode comparable to a compiled C function. The techniques described here could be applied to any language, but Python is popular in HEP for data exploration.

We should also note that the code transformation technique described here is similar to that of Mattis {\it et al}.\cite{columnarobjects}, though we statically transform and compile Python functions, whereas Mattis {\it et al}.\ implemented object proxies in PyPy and let PyPy's tracing JIT compiler dynamically optimize them. This technique can be viewed as a general alternative to object deserialization: when faced with user code that expects objects but the data are in another form, one could either transform the data to fit the code's expectation (deserialization) or transform the code to fit the data. When manipulating code, the data format is only interpreted once at compile-time; when manipulating data, the format is interpreted once per object, with extra memory allocations and copying, so code transformation is preferable when possible. Code transformation for columnar data is becoming a common technique in databases and search engines\cite{searchengine}, and we apply it to HEP data with a view toward building it into a HEP database, comparing its performance to analysis on materialized C++ objects in ROOT.

\section{Data Representation}

\subsection{PLUR Type System}
\label{type-system}

We begin by describing the scope of data types we are considering. To simplify the transformations, we restrict this set as much as possible while still being useful. In particular, the data types described here only encode data; they don't determine how data are used, such as which functions can be legally called on them. However, an interpretation layer can be overlaid on this representation without affecting the format or code transformations.

The set of possible types is generated by four constructors:
\begin{itemize}
\item {\bf Primitive:} fixed byte-width booleans, integers, floating point numbers, and characters. Even fixed-size matrices of numbers could be considered primitives--- the important point is that the width is known to the compiler.
\item {\bf List:} arbitrary-length, ordered collections of other types. Each instance may have a different width, including empty lists, and the list may contain any other type, including nested lists and Lists of Records. All objects in the list must have the same type (homogeneous).
\item {\bf Union:} an object that may be one of several types (``sum types'' in type theory). Each instance can have a different type, but its type must be chosen from a predetermined list. This provides more flexibility than class inheritance (e.g.\ {\tt Particles} that may be {\tt Muons} or {\tt Jets}), but less than dynamically typed Python.
\item {\bf Record:} containers mapping field names to types (``product types'' in type theory). Each instance must contain all fields, like a class in C++ or Python.
\end{itemize}
We call this type system PLUR, an acronym of the four constructors. A complete type schema in PLUR is a tree of Lists, Unions, and Records whose leaves are Primitive types.

One thing to notice is that this system does not allow for recursively defined types. For instance, one cannot make a {\tt Tree} Record containing {\tt Trees}. Thus, all data structures have a finite maximum depth determined by the schema. In practice, this is not a limitation, as trees and even arbitrary graphs can be (and routinely are) built in HEP by pointing to members of other subcollections with list indices.

Another thing to notice is that we have chosen not to require names for Records, as classes in C++ and Python must be named. This is to allow for more flexible type-checking, in which the structure of a Record (a minimum set of field names and types) is sufficient to determine if it can be used in a function. For example, a {\tt mass} function only needs to verify that the two particles it is given have {\tt pt}, {\tt eta}, and {\tt phi} fields, instead of having to introduce type annotations or explicit polymorphism into Python code.

Stronger type safety can be applied by overlaying this type system with names and adding dispatch rules. For instance, a List$\langle${\tt byte}$\rangle$ can be distinguished from a string of text with a name like {\tt UTF8String}. Functions like {\tt capitalize} would be applicable to strings in a way that they are not applicable to List$\langle${\tt byte}$\rangle$, and functions like {\tt len} could return the number of variable-width Unicode characters, rather than the number of raw bytes. However, these details are not important for the columnar representation: {\tt UTF8Strings} and List$\langle${\tt bytes}$\rangle$ are stored and accessed the same way, so the PLUR type system does not make a distinction. Similarly, unordered collections like sets and key-value mappings are just Lists at the storage level.

\subsection{OAMap: Objects to Arrays}

Any data that can be described by a PLUR schema can be represented in columnar arrays. This mapping from objects to arrays is analogous to Object Relational Mapping (ORM) of databases, so we call it Object Array Mapping (OAM).

HEP data in ROOT files are encoded in an OAM, though ROOT's encoding transforms any object with a C++ type into arrays. The C++ type system would be a large project to convert from data transformation rules into code transformation rules, so we limit our discussion to the PLUR type system. The data transformation rules we describe below happen to be very similar to ROOT's and also a subest of Apache Arrow\cite{arrow-layout}. We call it OAMap, and provide an implementation\cite{oamap} on GitHub. In our tests, we convert ROOT data into OAMap on-the-fly.

OAMap does not specify a storage mechanism: any means of storing arrays in a namespace may be used. This could be a Python dict of Numpy arrays, an HDF5 file, a filesystem directory of raw files, or a distributed object store. Only a PLUR schema is required to interpret the data, and this schema can even be encoded as a naming convention in the names of the arrays, eliminating the need for type metadata. The following set of rules transform an object of type $T$ with name $N$ into a set of named arrays.

{\bf If $T$ is a Primitive}, append the primitive value to an array named $N$, creating it if it doesn't yet exist.

{\bf If $T$ is a List} with contained type $T'$ and length $\ell$, find an array named $N + \mbox{``-Lo''}$ (list offset). If it does not yet exist, create it with a single element $0$. Then select the last element $e$ from this array and append $\ell + e$.

Next, iterate through each item in the List and apply the rule for $T'$ with name $N + \mbox{``-Ld''}$ (list data).

{\bf If $T$ is a Union} with possible types $T_1, \ldots, T_n$ and the value has actual type $T_t$ (where $t \in [1, n]$), find or create an array named $N + \mbox{``-Ut''}$ (union tag) and append $t$.

Next, follow the rule for name $N + \mbox{``-Ud''} + t$ (union data~$t$) and type $T_t$.

{\bf If $T$ is a Record} with field names and field types $(N_1, T_1), \ldots (N_n, T_n)$, follow the rule for each pair $N_f$, $T_f$ (where $f \in [1, n]$), using $N + \mbox{``-R\_''} + N_f$ (record field $N_f$) as a name and $T_f$ as a type.

A Record does not generate any arrays to represent its structure (as Lists and Unions do); the connection between fields in an array is entirely encoded in the PLUR schema or naming convention. To ensure that array names can be properly parsed, field names must not contain the character ``-'' (or a different delimiter should be chosen).

One must be sure to include empty/trivial arrays for types that were not touched due to missing data (Lists that are all empty at a given level or Union type possibilities that never occur in the data) or make the reading procedure insensitive to missing arrays. A simple way to include all arrays is to create them with a first pass over the type schema and only append to them in the pass over data.

\subsection{OAMap: Arrays to Objects}

The procedure described above losslessly stores the PLUR schema in array names and the object data in arrays. To demonstrate this, we describe an algorithm below that would materialize objects from arrays. As explained in the introduction, we prefer code transformation to object materialization, but it is important to prove that the transformation is indeed lossless.

First, select arrays whose names begin with the prefix $N$ and pop the prefix from their names. For each of these arrays $a$, create an index $i_a$ whose initial value is 0. Then recursively apply the following rules.

{\bf If only one array exists and its name is the empty string}, then the type is a Primitive. Return the value at index $i_a$ and increment $i_a$ by~1.

{\bf If one array name begins with ``-Lo'' and all others begin with ``-Ld''}, then the type is a List. Take $a[i_a + 1] - a[i_a]$ as the length $\ell$ of the List and increment $i_a$ by~1 for the array $a$ that begins with ``-Lo''.

Pop the ``-Ld'' from the beginning of all other array names and apply the rule for that set of arrays $\ell$ times to fill the List's contents.

{\bf If one array name begins with ``-Ut'', and all others begin with ``-Ud''}, then the type is a Union.

Take $a[i_a]$ as the tag $t$ for the datum and increment $i_a$ by~1 for the array $a$ that begins with ``-Ut''.

Pop the $\mbox{``-Ud''} + t$ from the beginning of all other array names associated with tag $t$ and apply the rule for that subset.

{\bf If all array names begin with ``-R\_''}, then the type is a Record. Partition the set of arrays by field name $N_f$, pop the $\mbox{``-R\_''} + N_f$ from the beginning of the array names, and apply the rule separately for each partition to fill each field.

If any other configuration of arrays and names is encountered, the arrays are malformed. If indices go beyond the lengths of the arrays or do not perfectly end on the last element of each array, then the arrays are malformed.

\subsection{Random Access and Redundancy}
\label{random-access-and-redundancy}

The reason that List structures are represented by data offsets (arrays whose names end in ``-Lo''), rather than List lengths, is to permit random access. For instance, if we had a List$\langle$List$\langle${\tt float}$\rangle\rangle$ named ``x'' and we wanted the {\tt float} at index $(i, j)$, we would compute
\[ \mbox{x-Lo-Ld{\tt [}x-Lo-Lo{\tt [}x-Lo{\tt [}} 0 \mbox{\tt ]} + i \mbox{\tt ]} + j \mbox{\tt ]} \]
At each level, the contents of the ``-Lo'' array are the starting indices of the next-deeper structure.

In fact, if we want to reconstruct only one object in a List, we simply apply the arrays-to-objects algorithm for that List's ``-Ld'' arrays with all indices starting at $i_a = i$.


The Union structure, as described so far, is not random accessible. Data arrays for each type possibility $T_t$ are only filled each time a value of that type is encountered, which cannot be every time for every type. Each type possibility must be indexed by a different offset, but these offsets can all be packed into the same array becuase exactly one type must be encountered per instance. The arrays-to-objects algorithm described in the previous subsection avoids this issue by walking over the data sequentially.

To make Unions random accessible, we need to add a union offset array ``-Uo'', which can be generated from the tag array ``-Ut'' by
\begin{algorithmic}
\vspace{0.15 cm}
\STATE $i_t \coloneqq 0$ {\bf for all} $t \in [1, n]$

\vspace{0.15 cm}
\FOR{$i \coloneqq 0$ {\bf until} length of x-Ut}
\STATE $t \coloneqq$ \mbox{x-Ut}{\tt [}$i${\tt ]}
\STATE \mbox{x-Uo}{\tt [}$i${\tt ]} $\coloneqq i_t$
\STATE $i_t \coloneqq i_t + 1$
\ENDFOR
\end{algorithmic}

Now we can access Union objects randomly: to reconstruct a Union at index $i$, we find the tag $t$ at x-Ut{\tt [}$i${\tt ]} and follow the arrays-to-objects algorithm on the set of arrays named with the corresponding ``-Ud$t$'', all starting at index $i_t$ given by $\mbox{x-Uo{\tt [}}i\mbox{\tt]}$. Arrow calls this a ``dense union.''

\subsection{Relationship to Apache Arrow and ROOT}

We chose to implement a subset of the Arrow AOM to increase the usefulness of our tools beyond HEP. Arrow unifies in-memory data frames across a variety of Big Data platforms\cite{arrow}, so a code transformation tool that assumes this encoding may be directly applied to data from these platforms. The version of OAMap tested here lacks Arrow's nullable types, but it is being extended to cover these cases. OAMap generalizes beyond Arrow in that arrays may be provided on demand for large, out-of-memory datasets.

ROOT's encoding differs from OAMap in that some List offsets are represented as byte offsets, which can be converted to object offsets by an affine transformation, and in some cases as List lengths, which must be summed. Also, OAM is optional in ROOT and is only carried out one List level deep, but most HEP data are presented in this form anyway. Different C++ types like STL vectors and arrays are encoded differently, so we normalize all list-like types to PLUR Lists in our on-the-fly conversion.



\section{Code Transformation}

OAMap's most important feature is that the structure of any type can be expressed by a single integer: the index that would be used to start the arrays-to-objects algorithm. A type and a name prefix uniquely specifies a set of arrays, so the only values required at runtime are the locations of instance data within those arrays.

Only one index is required because:
\begin{itemize}
\item a Primitive value is located at one index in its array;
\item a List's length can be computed from one index \mbox{($a[i_a + 1] - a[i_a]$)} and all of its contents are derived from the value of its offset array at that index;
\item a Union is specified by a tag from its tag array and an offset from its offset array, which occur at the same index. All of the contents are derived from the offset;
\item a Record is just a bundle of fields with no structure array. However, the contents of all fields in a Record object start at the same index.
\end{itemize}

Therefore, any function that operates on Primitive, List, Union, and Record objects, no matter how complicated, can be replaced with a function that operates on integer indices. Replacing each instance of a PLUR-typed object with its index and all functions and code constructs operating on objects with the equivalent operations for indices transforms the code to match the data encoding, rather than shaping the data to match the object-oriented vocabulary of the code.

To see this as an alternative to deserialization, consider an extreme scenario in which all arrays are in a binary file on disk that has been memory-mapped to look like an array. Wherever the source code would have required a structured object, the compiled code operates directly on the raw bytes on disk. No additional representation of the data is created.

This technique can be applied to code written in any language, but all of the explicit examples are applied to Python code because this is our chosen query language.

\subsection{General Strategy}

Unlike SQL, functional and procedural programming languages can assign and possibly reassign variables, extract substructure as new variables, loop over substructure, and pass objects to other functions, where they are identified with new names. It is not sufficient to only transform symbols that coincide with object names, since parts of the data structure may be spread among user-defined variables. To avoid missing code that needs to be transformed, we must track the PLUR type of all symbols in the function.

We therefore need a typed AST, in which PLUR types and array names are associated with all expression nodes that hold non-Primitive PLUR type. (Knowledge of other types is not necessary but not harmful.) These types and names must be propagated from symbols to expressions through assignment operators so that the correct array references may be injected into the code. We performed code transformation and type propagation in a single sweep.

There are several constraints on the code to be transformed. It cannot create or change PLUR-typed objects, which is reasonable for a query language in which the input dataset and auxiliary inputs are immutable. The analysis function may call external functions, but the AST of those functions must either be accessible so they can be transformed, too, or they must accept only Primitive or non-PLUR arguments. Functions cannot be passed as objects and then called on PLUR types, since an unknown function can't be statically transformed. A variable name can't have different PLUR types in the same scope, which is a normal constraint for statically typed code, but unusual for dynamically typed Python. All of these are restricted by the Numba compiler as well, and since we pass our transformed functions to Numba, they do not represent {\it additional} constraints. Our code transformation process may be viewed as extending Numba from arrays and simple Python types to include any immutable, PLUR-typed object.

The following AST nodes must be transformed:
\begin{itemize}
\item {\bf symbol reference}, which might have PLUR type;
\item {\bf assignment}, which can pass PLUR type from an expression to a new symbol;
\item {\bf list subscript} (square brackets in most languages), which might slice or extract from a PLUR List;
\item {\bf attribute subscript} (dot in most languages), which might extract a field from a PLUR Record;
\item {\bf function calls}, which can pass PLUR types to a new function, widening the scope of the transformation sweep to include that function, and may return a new PLUR type/prefix that must be tracked;
\item {\bf for loops} which might iterate over a PLUR List;
\item {\bf special functions}, like {\tt len} (List length) and {\tt isinstance} (check type) in Python, which have to be handled in special ways when called on PLUR types.
\end{itemize}

Any use of a PLUR-typed object that isn't specially handled must be treated as an error to avoid incorrect code, since these objects are replaced by plain integers at runtime. These errors would appear to the user as compilation errors, with line numbers and meaningful error messages.

\subsection{Required Transformations}

{\bf Symbol references} are leaves of the AST and therefore first to be transformed in the recursive walk. In the original function, an identifier that refers to a PLUR-typed object becomes an identifier for its integer index, so a symbol reference becomes array extraction.

For Primitive and List types:
\begin{center}
{\tt x} (referring to object) $\to$ {\tt array[x]} ({\tt x} is index)
\end{center}
where {\tt array} is the array associated with a Primitive or the offset array associated with a List.

Union objects should be immediately replaced with an object of specific type. Since that type is not known during code transformation, every branch must be followed, and subsequent code transformations must be predicated on the runtime tag value. The symbol table must therefore be branchable, a list of possible symbol tables that multiplies as unions are encountered. Union types lengthen the compilation process, but have minimal impact on runtime (an extra integer check). Branching is tamped by any {\tt isinstance} checks written by the user (see below).

References to Record symbols do not require any transformation, though they are reinterpreted as indices that would be passed to their fields when subscripting (see below).

{\bf Assignment} merely passes the PLUR type and associated array names to a new symbol. In the type-inference pass, this means adding the type information to a symbol table.

Assignment is more complex if one tries to handle pattern matching, such as Python's tuple unpacking (see ``Flourishes'' below).

{\bf List subscripts} replace a List index with a Primitive value if it is a List of Primitives or the index of the next substructure down if it is any other type. This can be performed in two steps: (1) transform
\begin{center}
{\tt x[i]} $\to$ {\tt offset[x + i]}
\end{center}
and then (2) transform the result of this as though it were a symbol reference (rule described above) using the List's contained type.

An attempt to subscript any PLUR type other than a List is an error. The above only works if the index resolves to integer type, not (for example) a Python {\tt slice}. Handling slices would be considered a flourish.

{\bf Attribute subscripts} extract a field from a Record by name. In most languages, the field names are syntactically required to be a string known at compile-time with constraints on the characters.

We do this transformation in two steps, like the list subscript above: (1) transform
\begin{center}
{\tt x.fieldname} $\to$ {\tt x}
\end{center}
and then (2) transform the result of this as though it were a symbol reference (rule described above) using the selected field type.

An attempt to subscript any PLUR type other than a Record is an error.

{\bf Function calls} may include non-Primitive PLUR types in their arguments or not. If a function does not take any non-Primitive PLUR types as arguments, it can be left as-is.

If it references non-PLUR types, then we must obtain the AST for that function and propagate PLUR types through it, starting from the argument types. If the function has previously been transformed with the same types, we may reuse the previously transformed function (as long as we pass array names as its new arguments). If it was transformed with different types, we must generate a new transformed copy of the function, propagating the new PLUR types through it. We are effectively treating the function as though it were type-parameterized in every argument, the way that Julia\cite{julia} does with functions that don't have type annotations.

For example, a {\tt mass} function that takes two arguments, {\tt particle1} and {\tt particle2}, would be transformed twice if called with ({\tt Muon}, {\tt Muon}) types at one site and ({\tt Muon}, {\tt Jet}) at another. The second transformation verifies that the {\tt Jet} Record has all the required fields to calculate a mass.

If a function is recursive, this process would not terminate without return-type hints (e.g.\ for any legal inputs, {\tt mass} returns a floating-point number). Refusing to transform recursive functions would not overly restrict HEP analysis functions, though Python 3's type annotations may be helpful for such cases.

A function may return PLUR type; the transformed function either returns a Primitive or an integer index that must be propagated from the call point in the original function.

{\bf For loops} may iterate over a PLUR List. The transformed List is just an integer index for the offset array, so it must be replaced by an iterator over offsets:
\begin{center}
{\tt x} $\to$ {\tt range(array[x], array[x + 1])}
\end{center}
The loop variable is now an integer representing indices into the List's contents, as desired.

{\bf Special functions} that return meaningful data about non-Primitive PLUR types must be handled on a case-by-case basis. As with any function call, if a non-Primitive PLUR type is an argument, the function's identity must be known during the transformation sweep. Below are two important cases.

{\bf len}: (get length of List) must return the length when given a PLUR List, which is computed as
\begin{center}
{\tt len(x)} $\to$ {\tt off[x + 1] - off[x]}
\end{center}
{\it without} transforming the argument {\tt x} (unlike other function calls, which operate on transformed arguments).

{\bf isinstance}: (check type) must return true if the argument has a given type, false otherwise. If the symbol table is branched, some symbol tables may be eliminated in scopes guarded by an {\tt isinstance} check (``{\tt and}'' and ``{\tt if}''). This is the primary way a user might benefit from a Union type.

The type or types to check would have to be known as literal names, not computed references, during the transformation sweep. If the argument to check has List or Record type, the whole expression can be statically replaced with a literal {\tt True} or {\tt False}, depending on whether the desired types are in the set of types to check. If it is a Union, it must be replaced with a tag check.

\subsection{Flourishes}

The implementation of the code transformation is open-ended: one can provide more or less support for PLUR-typed objects, as well as optimizations. The only invariant is that the transformed code must either work exactly as object-oriented code would or fail to be transformed (compilation error).

\subsubsection{List Overflows}

One special case requires special attention: what to do about List index overflows? In the minimal transformation described above, PLUR List overflows behave like C array overflows, in that they return undefined results. The errors are less obvious and therefore more dangerous than typical C array overflows because PLUR values just beyond a List's boundary belong to the same attribute in the next List, so they probably have the right scale and distribution, subtly biasing analysis results.

A simple way to eliminate mistakes like this is to add a range check to the transformed code. Indices that fail the range check should raise a runtime exception. It would be correct, but many of these range checks would be unnecessary and would slow down processing. For instance, the list subscript might be in the body of an {\tt if} statement where the user did range-checking manually, or it could be in a {\tt for} loop bounded by the list length (a common case).

A conservative approach would be to apply range checking by default and remove it from unnecessary cases as they are identified. The performance tests in the last section have no range checks, but the OAMap library has range checking at the time of writing.


\subsubsection{Pythonic Indices}

Negative index handling in Python is a user-familiarity enhancement, in which negative values start counting from the end of the list. Without this enhancement, negative indices would be caught by a range check, so it is not strictly required. Without a range check, it is extremely dangerous, as users would get subtly wrong results by assuming normal Pythonic behavior. (Pythonic indices are currently implemented in OAMap.)

\subsubsection{Eliminating Zero-Lookups}

As an optimization, one can statically identify array lookups that always return zero: the first element of every list offset array is zero, and the outermost list offset array is always evaluated at its first element. Without explicitly removing it, this unnecessary code would be executed at every step in an iteration over a List. (PLUR-unaware compilers cannot remove it.)

\subsubsection{For loop flattening}

Related to the above, nested {\tt for} loops that exhaustively walk over List contents, such as
\begin{center}
\begin{minipage}{0.8\linewidth}
\begin{Verbatim}[commandchars=\\\{\}]
\PY{o+ow}{for} \PY{n}{inner} \PY{o+ow}{in} \PY{n}{outer}:
  \PY{o+ow}{for} \PY{n}{x} \PY{o+ow}{in} \PY{n}{inner}:
    \PY{n}{do\PYZus{}something}(\PY{n}{x})
\end{Verbatim}

\end{minipage}
\end{center}
can be collapsed into a single {\tt for} loop over the innermost contents ({\tt x}). The innermost data are stored contiguously, so a single loop would suffice, and it is much more likely to be vectorized by the compiler. (A PLUR-unaware compiler cannot make this optimization because it does not know that list offset arrays are monotonically increasing.)

\subsubsection{Fixed-size Arrays and Matrices}

PLUR Lists can have any length, which includes a constant length, at the expense of redundant offset arrays with linearly increasing contents.

It would be possible to add the concept of a fixed-length List to the type system and propagate its implementation through OAMap and the code transformation rules. However, an important special case in which the fixed-length dimensions directly contain Primitives is available for free by accepting Numpy arrays with multidimensional {\tt shape} parameters as Primitives.

\subsubsection{Type Constructors}

As stated above, passing a non-Primitive PLUR object to any unrecognized function is a compilation error. This especially includes constructors for dynamic objects like Python lists and dictionaries, since we cannot statically track where they are used. However, one may wish to allow some immutable objects to contain PLUR objects, such as Python tuples to allow Python tuple-unpacking in assignments. We then become obliged, however, to track these objects as containing PLUR types.

\subsubsection{Pattern Matching}

Some languages make extensive use of typed pattern matching; PLUR types would need to be tracked through syntactical structures such as these. In Python, tuple-unpacking is the most obvious instance of pattern-matching, and type inference through it is possible (implemented in the version of OAMap used in these tests).

\subsubsection{Equality and Order}

Since PLUR types describe inert data (as opposed to functions or active elements like file handles), it would be reasonable to define value-based equality and possibly an ordering, effectively treating comparison operators like ``{\tt ==}'' and ``{\tt <}'' as special functions (as well as ``{\tt sorted}'').

Reference-based equality, expressed in Python as an ``{\tt is}'' operator, is easiest to implement. For objects typed at identical nodes of the PLUR schema, references are the same if their indices are equal. At compile-time, this is a schema comparison, and at runtime, we replace ``{\tt is}'' with ``{\tt ==}''.

\subsubsection{Fallback to Object Materialization}

One way to support particularly difficult cases, such as external functions without code transformation, is by materializing objects using the arrays-to-objects algorithm. This gives up on zero-copy efficiency, but it may be worthwhile to mix fast code with expressive code. Then, rather than new features allowing specific cases of user code to run at all, new features would allow the user code to run faster.

\subsubsection{Function dispatch}

As described in Section~\ref{type-system}, the PLUR type system only distinguishes between storage types. If the same storage type with different names should be dispatched to different functions or different versions of a function, that would be handled in code transformation. As with special functions ({\tt len} and {\tt isinstance}), the identities of these functions must be known at compile-time.

\section{Implementation}

We are developing the OAMap toolkit\cite{oamap} as the execution engine of a HEP database/query system, but it is also usable on its own. This toolkit includes the code transformation routine outlined above, which is intended for high-throughput processing, and proxy classes for low-latency exploration on the Python commandline. The proxies are Python classes that yield data on demand, using Python's {\tt property}, {\tt \_\_getitem\_\_}, and {\tt \_\_getattr\_\_} to emulate static members by fetching data (from memory, disk, or network) as necessary. These proxies most clostly resemble the work of Mattis {\it et al}.\ in PyPy\cite{columnarobjects}, except that PyPy is JIT compiled on the fly, making the proxy and code transformation approaches equivalent. However, Numpy and CPython provide access to much-needed scientific libraries.

The use of proxies would also be equivalent to code transformation in a fully compiled language like C++. The propagation of PLUR type would be performed by the C++ compiler, rather than manually through a Python AST. However, C++ would not be a convenient query language, and type-level programming in C++ is not as powerful as direct AST manipulation. (For example, it might not be possible to collapse {\tt for} loops based on our knowledge that list offset arrays are monotonically increasing.) Julia would be an excellent compromise, as it is a high-level language without manual memory management, it automatically JIT compiles, and provides access to the Julia AST, but C++ and Python are much more commonly used in HEP.

Nearly all HEP data are stored in ROOT files, so we must be able to read this format efficiently. We have implemented a modification to ROOT, called BulkIO, that allows us to skip ROOT's usual object materialization methods ({\tt TTree::GetEntry} and {\tt TBranch::GetEntry}). These updates are scheduled to become part of the base ROOT distribution in ROOT version 6.14.

Although the performance studies in the next section use ROOT with the BulkIO enhancements, this access method is also available as a pure Python package called uproot\cite{uproot}.

\subsection{Performance Studies}

As an execution engine, the transformed code must be convenient and fast. We chose Python over C++ for convenience and must not pay for that choice in performance. For this reason, we have been performance-testing OAMap throughout its development, informally comparing against ``bare metal'' speeds of one-off C programs. We always use Numba with ``{\tt nopython=True},'' so the code it compiles with LLVM is almost perfectly equivalent to a C program compiled with Clang. Our informal tests bore this equivalence, but were unrealistically sourced with random data.

Real-world uses of OAMap would be sourced with HEP data in ROOT files (or a database equivalent). The most relevant comparison then is between a C++ analysis function in ROOT, from ROOT data through object materialization, and an OAMap-transformed Python function, from ROOT data but accessed as in-place arrays. We used the BulkIO enhancements to stream data into Numpy arrays and OAMap to transform and then compile the same analysis functions. C++ and Python versions of the analysis functions are listed in Figure~\ref{four-functions}.

\begin{figure}
\scriptsize

\noindent\begin{minipage}{\textwidth}
\begin{minipage}[c][1.8cm][t]{0.22\linewidth}
\underline{{\bf max p$_{\mbox{\tiny T}}$} in Python}
\input{minted-2.tex}
\end{minipage}
\begin{minipage}[c][1.8cm][t]{0.25\linewidth}
\underline{{\bf max p$_{\mbox{\tiny T}}$} in C++}
\input{minted-3.tex}
\end{minipage}

\vspace{0.25 cm}
\begin{minipage}[c][2.6cm][t]{0.22\linewidth}
\underline{{\bf eta of best by p$_{\mbox{\tiny T}}$} in Python}
\input{minted-4.tex}
\end{minipage}
\begin{minipage}[c][2.6cm][t]{0.25\linewidth}
\underline{{\bf eta of best by p$_{\mbox{\tiny T}}$} in C++}
\input{minted-5.tex}
\end{minipage}%

\vspace{0.25 cm}
\begin{minipage}[c][3.2cm][t]{0.22\linewidth}
\underline{{\bf mass of pairs} in Python}
\input{minted-6.tex}
\end{minipage}
\begin{minipage}[c][3.2cm][t]{0.25\linewidth}
\underline{{\bf mass of pairs} in C++}
\input{minted-7.tex}
\end{minipage}%

\vspace{0.25 cm}
\begin{minipage}[c][2.4cm][t]{0.22\linewidth}
\underline{{\bf p$_{\mbox{\tiny T}}$ sum of pairs} in Python}
\input{minted-8.tex}
\end{minipage}
\begin{minipage}[c][2.4cm][t]{0.25\linewidth}
\underline{{\bf p$_{\mbox{\tiny T}}$ sum of pairs} in C++}
\input{minted-9.tex}
\end{minipage}%
\end{minipage}

\caption{\label{four-functions} Sample analysis functions in Python (before code transformation) and object-oriented C++, showing only the body of the loop over events.}
\end{figure}

The first, ``max p$_{\mbox{\scriptsize T}}$,'' is an example of a query that would be difficult but not impossible in SQL. Instead of exploding a muons table and grouping by a unique {\tt eventId}, we keep a running maximum that resets in each event.

The second, ``eta of best by p$_{\mbox{\scriptsize T}}$,'' is an extension of that idea: we select a muon by maximizing {\tt pt} and then plot its {\tt eta}. This is even more awkward in SQL, but very common in HEP.

In the listing, note that {\tt best}, which is a muon object, is initialized as $-1$ in Python and {\tt nullptr} in C++. At the time of testing, the code transformer had no concept of a nullable PLUR type, though OAMap has this feature now. Instead of initializing the object as $-1$ and checking for that value as a negative index, we can now write it more naturally as {\tt None} and assigning the same variable as a muon Record and a number would be a compile-time error.

The third function, ``mass of pairs,'' would require a full outer join in SQL but is a nested for loop in Python and C++, carefully indexed to avoid duplication. This kind of ``unique pairs'' loop is very common (often with a selection, e.g.\ requiring opposite-sign charges), and the {\tt mass} formula is one of the most frequently executed in HEP.

The fourth, ``p$_{\mbox{\scriptsize T}}$ sum of pairs,'' is a diagnostic of the third. As we will show below, the mass calculation is the slowest of the sample functions, and it was unclear whether the nested indexing was responsible or the complex formula. This function has the same nesting structure but a simpler formula (one that is occasionally useful in HEP).

We ran each analysis function on a 5.4 million event dataset of simulated Drell-Yan collisions (a realistic physics sample, one of about a dozen that might be involved in a real HEP analysis). The tests were performed on an {\tt i2.xlarge} instance on Amazon Web Services, which features a large, fast SSD (though in the end, we opted for tests with prewarmed cache).

To avoid decompression and/or physical device access as a dominant contributor to these tests, we prepared an uncompressed ROOT file and loaded it into Linux page cache with vmtouch\cite{vmtouch}. This lets us see the differences due to other factors more clearly.

Both tests were single-threaded and had plenty of working space memory. The benefits of parallelization are beyond the scope of this paper, and also factorize from our single-threaded tests. If we double the single-threaded speed of this embarrassingly parallel problem, we double the speed of a parallelized version (unless that parallel processor is swamped with overhead).

\begin{figure}[!t]
\centering
\includegraphics[width=\linewidth]{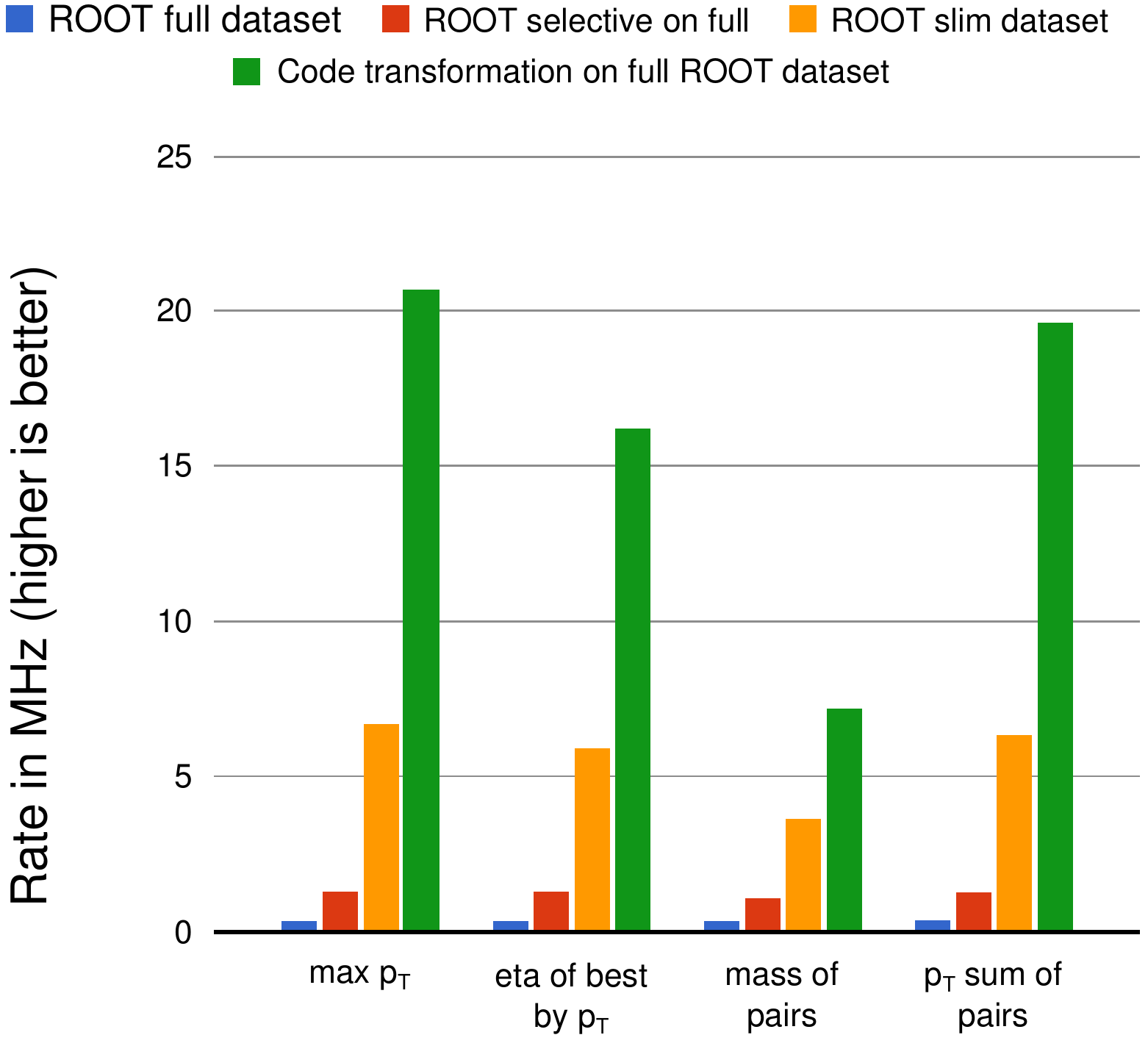}
\caption{Event processing rates, including read and execute times, for 5.4 million events in the test sample. Grouped bars indicate different analysis functions and colors indicate different workflows. See the text for details.}
\label{root-and-plur}
\end{figure}

The results of the study are shown in Figure~\ref{root-and-plur}. The shorthand ``ROOT'' signifies a conventional ROOT workflow with object materialization and C++, and ``code transformation'' is our workflow with BulkIO, no object materialization, and transformed Python code.

The four colors signify variants on these workflows. ``ROOT full dataset'' means we let ROOT fill all 42 attributes of the Muon objects, which is clearly unnecessary for our functions. The event processing rate for this case is 0.4~MHz, regardless of the content of the function. Reading and filling the objects overwhelms all other factors. (This case is only included for completeness.)

``ROOT selective on full'' uses ROOT's opt-in mechanism to avoid filling all attributes in the objects, but still used the 42-field object definitions and original data file. The event processing rate is 1.29~MHz, regardless of the function. Handling all of the attributes still dominates.

``ROOT slim dataset'' performs the same selective read on a specially prepared dataset and object definition that has only 3 fields: {\tt pt}, {\tt eta}, and {\tt phi}. We now see different rates for the four functions: 6.68, 5.96, 3.56, and 6.34~MHz.

``Code transformation on full ROOT dataset'' is the only test in this batch that uses transformed Python code. It accesses the full dataset; the slim dataset yields similar results because this method is unconcerned with unused columns. The four functions are processed at 17.9, 12.1, 6.09, and 17.2~MHz, respectively, considerably faster than object materialization.

The slowest of the four functions is ``mass of pairs.'' Unlike the first two functions, this involves a doubly nested loop over muons to find distinct pairs, as well as a much more complex formula. The fourth function has the same loop structure without the complex formula, and it is as fast as the single loops. Upon further investigation, we found that the trigonometric and hyperbolic cosines account for the majority of the time spent in this function.

The factor that the ROOT tests and OAMap test had in common is that they both extracted data from ROOT files and used part of the ROOT framework to read them. OAMap can operate on any arrays, regardless of whether they came from ROOT, so we performed another test in which we extracted all the data as Numpy arrays in memory. The ROOT files were in memory, too, because we prewarmed the Linux cache with vmtouch, but some processing is still required to seek to the relevant parts of the file to expose the arrays.

\begin{figure}[!t]
\centering
\includegraphics[width=\linewidth]{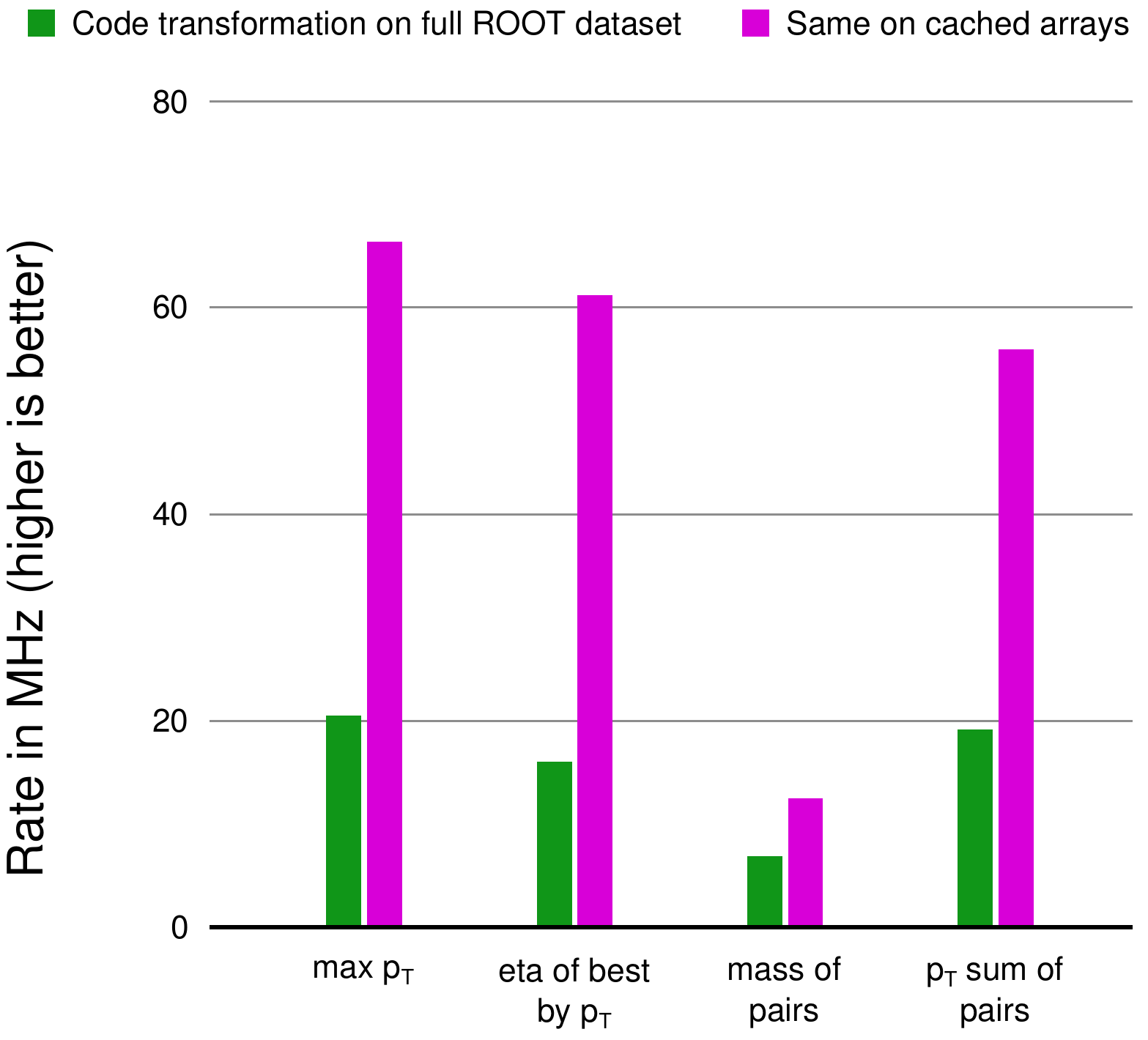}
\caption{Event processing rates for the same four analysis functions, but only for PLUR workflows. Colors represent data sources: full ROOT dataset and Numpy arrays in memory.}
\label{physical-media}
\end{figure}

We compare the OAMap result (copied from the previous plot on the new scale) from ROOT and from raw arrays in Figure~\ref{physical-media} for two reasons: (1) it shows just how much room there is between these execution rates and the data-access rate, since somewhat more complex functions will be slower and we want to know when to expect it to be a bottleneck, and (2) because a HEP database system would conceivably cache frequently accessed arrays in memory, and this shows the difference between a cache-hit and a cache-miss. The effect of the slow mass calculation is dramatic: ``mass of pairs'' runs at 12.8~MHz while ``p$_{\mbox{\scriptsize T}}$ sum of pairs'' runs at 56.2~MHz.

\section{Conclusions}

We have demonstrated that it is possible to transform code to meet a data format, rather than deserializing data to fit the code's expectations, as long as JIT compilation is available. We have also demonstrated that, once transformed and compiled, an idiomatic Python analysis function outperforms the same function written in idiomatic C++ with object materialization.

This should not be taken as a claim that Python is faster than C++, just as SQL is not faster than C++, but that we have applied one of the tools used by databases to accelerate queries to a general-purpose language. The compilation of Python by Numba provides parity between a restricted subset of Python (statically typable, no first-class functions) and C++, and the code transformation avoids the cost of object materialization. In principle, the same speedup could be achieved in C++ or Julia with the appropriate proxy classes.

Finally, the usefulness of this technique is not limited to HEP. The need for complex loop dependencies can hardly be HEP-specific\footnote{As an indication of this need, Netflix proposed changes to Spark to allow higher-order functions in SparkSQL, which would transform subcollections without a full explode-and-join: {\tt https://issues.apache.org/jira/browse/SPARK-22231}}. Although many industries and fields of academia are currently using SQL or languages similar to SQL for data analysis, how much is being left unexplored because the tools are not suited for the task?

It is our hope that this technique finds application in many different fields, just as the techniques of database-style analysis have inspired our work in HEP.

\section*{Acknowledgments}

This work was supported by the National Science Foundation under Grants No.\ 1450377 and 1450323. The authors wish to thank Philippe Canal for his expert help in understanding the ROOT file format and ROOT I/O subsystem.

\bibliographystyle{IEEEtran}
\bibliography{IEEEabrv,mybibliography}

\begin{thebibliography}{10}
\providecommand{\url}[1]{#1}
\csname url@samestyle\endcsname
\providecommand{\newblock}{\relax}
\providecommand{\bibinfo}[2]{#2}
\providecommand{\BIBentrySTDinterwordspacing}{\spaceskip=0pt\relax}
\providecommand{\BIBentryALTinterwordstretchfactor}{4}
\providecommand{\BIBentryALTinterwordspacing}{\spaceskip=\fontdimen2\font plus
\BIBentryALTinterwordstretchfactor\fontdimen3\font minus
  \fontdimen4\font\relax}
\providecommand{\BIBforeignlanguage}[2]{{%
\expandafter\ifx\csname l@#1\endcsname\relax
\typeout{** WARNING: IEEEtran.bst: No hyphenation pattern has been}%
\typeout{** loaded for the language `#1'. Using the pattern for}%
\typeout{** the default language instead.}%
\else
\language=\csname l@#1\endcsname
\fi
#2}}
\providecommand{\BIBdecl}{\relax}
\BIBdecl

\bibitem{mongodb}
E.~{Botoeva}, D.~{Calvanese}, B.~{Cogrel}, and G.~{Xiao}, ``{Expressivity and
  Complexity of MongoDB (Extended Version)},'' \emph{ArXiv e-prints}, Mar.
  2016.

\bibitem{drill}
\BIBentryALTinterwordspacing
M.~Hausenblas and J.~Nadeau, ``{Apache Drill: Interactive Ad-Hoc Analysis at
  Scale},'' \emph{Big Data}, vol. 1(2), pp. 100--104, 2013. [Online].
  Available: \url{https://doi.org/10.1089/big.2013.0011}
\BIBentrySTDinterwordspacing

\bibitem{spark}
\BIBentryALTinterwordspacing
M.~Zaharia, R.~S. Xin, P.~Wendell, T.~Das, M.~Armbrust, A.~Dave, X.~Meng,
  J.~Rosen, S.~Venkataraman, M.~J. Franklin, A.~Ghodsi, J.~Gonzalez,
  S.~Shenker, and I.~Stoica, ``{Apache Spark: A Unified Engine for Big Data
  Processing},'' \emph{Commun. ACM}, vol.~59, no.~11, pp. 56--65, Oct. 2016.
  [Online]. Available: \url{http://doi.acm.org/10.1145/2934664}
\BIBentrySTDinterwordspacing

\bibitem{sparksql}
\BIBentryALTinterwordspacing
M.~Armbrust, R.~S. Xin, C.~Lian, Y.~Huai, D.~Liu, J.~K. Bradley, X.~Meng,
  T.~Kaftan, M.~J. Franklin, A.~Ghodsi, and M.~Zaharia, ``{Spark SQL:
  Relational Data Processing in Spark},'' in \emph{Proceedings of the 2015 ACM
  SIGMOD International Conference on Management of Data}, ser. SIGMOD
  '15.\hskip 1em plus 0.5em minus 0.4em\relax New York, NY, USA: ACM, 2015, pp.
  1383--1394. [Online]. Available:
  \url{http://doi.acm.org/10.1145/2723372.2742797}
\BIBentrySTDinterwordspacing

\bibitem{h2o}
\BIBentryALTinterwordspacing
{The H2O.ai Team}. (2015) {``H2O: Scalable Machine Learning. Version
  3.1.0.99999''}. [Online]. Available: \url{http://www.h2o.ai}
\BIBentrySTDinterwordspacing

\bibitem{theano}
\BIBentryALTinterwordspacing
{Theano Development Team}, ``{Theano: A {Python} framework for fast computation
  of mathematical expressions},'' \emph{arXiv e-prints}, vol. abs/1605.02688,
  May 2016. [Online]. Available: \url{http://arxiv.org/abs/1605.02688}
\BIBentrySTDinterwordspacing

\bibitem{julia}
\BIBentryALTinterwordspacing
J.~Bezanson, S.~Karpinski, V.~B. Shah, and A.~Edelman, ``{Julia: {A} Fast
  Dynamic Language for Technical Computing},'' \emph{CoRR}, vol. abs/1209.5145,
  2012. [Online]. Available: \url{http://arxiv.org/abs/1209.5145}
\BIBentrySTDinterwordspacing

\bibitem{root}
R.~Brun and F.~Rademakers, ``{ROOT: An object oriented data analysis
  framework},'' \emph{Nucl. Instrum. Meth.}, vol. A389, pp. 81--86, 1997.

\bibitem{bulkio}
\BIBentryALTinterwordspacing
{Brian Bockelman, Zhe Zhang, and Jim Pivarski}, ``{ROOT-BulkIO and Numpy
  interface},'' \emph{Journal of Physics: Conference Series}, 2017. [Online].
  Available: \url{https://indico.cern.ch/event/567550/contributions/2627167/}
\BIBentrySTDinterwordspacing

\bibitem{numba}
\BIBentryALTinterwordspacing
S.~K. Lam, A.~Pitrou, and S.~Seibert, ``{Numba: A LLVM-based Python JIT
  Compiler},'' in \emph{Proceedings of the Second Workshop on the LLVM Compiler
  Infrastructure in HPC}, ser. LLVM '15.\hskip 1em plus 0.5em minus 0.4em\relax
  New York, NY, USA: ACM, 2015, pp. 7:1--7:6. [Online]. Available:
  \url{http://doi.acm.org/10.1145/2833157.2833162}
\BIBentrySTDinterwordspacing

\bibitem{columnarobjects}
\BIBentryALTinterwordspacing
T.~Mattis, J.~Henning, P.~Rein, R.~Hirschfeld, and M.~Appeltauer, ``{Columnar
  Objects: Improving the Performance of Analytical Applications},'' in
  \emph{2015 ACM International Symposium on New Ideas, New Paradigms, and
  Reflections on Programming and Software (Onward!)}, ser. Onward! 2015.\hskip
  1em plus 0.5em minus 0.4em\relax New York, NY, USA: ACM, 2015, pp. 197--210.
  [Online]. Available: \url{http://doi.acm.org/10.1145/2814228.2814230}
\BIBentrySTDinterwordspacing

\bibitem{searchengine}
X.-F. Jia, A.~Trotman, and J.~Holdsworth, ``{Fast Search Engine Vocabulary
  Lookup},'' 12 2011.

\bibitem{arrow-layout}
\BIBentryALTinterwordspacing
{The Apache Arrow team}. (2016) {``From the Apache Arrow wiki: Physical memory
  layout''}. [Online]. Available:
  \url{https://github.com/apache/arrow/blob/master/format/Layout.md}
\BIBentrySTDinterwordspacing

\bibitem{oamap}
\BIBentryALTinterwordspacing
{Jim Pivarski}. (2017) {``OAMap: Toolset for Computing Directly on
  Hierarchically Nested, Columnar Data''}. [Online]. Available:
  \url{https://github.com/diana-hep/oamap}
\BIBentrySTDinterwordspacing

\bibitem{arrow}
\BIBentryALTinterwordspacing
{The Apache Arrow team}. (2016) {``Apache Arrow: Powering Columnar In-Memory
  Analytics''}. [Online]. Available: \url{https://arrow.apache.org/}
\BIBentrySTDinterwordspacing

\bibitem{uproot}
\BIBentryALTinterwordspacing
{Jim Pivarski}. (2017) {``Uproot: Minimalist ROOT I/O in pure Python and
  Numpy.''}. [Online]. Available: \url{https://github.com/scikit-hep/uproot}
\BIBentrySTDinterwordspacing

\bibitem{vmtouch}
\BIBentryALTinterwordspacing
{Doug Hoyte}. (2012) {``vmtouch --- the Virtual Memory Toucher''}. [Online].
  Available: \url{https://hoytech.com/vmtouch/}
\BIBentrySTDinterwordspacing

\end{thebibliography}

\end{document}